\documentstyle[aaspp4,psfig,flushrt]{article}

\newcommand{\minusone}{$^{-1}$}
\newcommand{\LOH}{$L_{OH}$ }
\newcommand{\LFIR}{$L_{FIR}$ }

\righthead{OH Megamasers I:  Preliminary Results}
\lefthead{Darling \& Giovanelli}

\begin{document}
\title{A Search for OH Megamasers at $z > 0.1$. \ \ I.\  Preliminary Results}
\author{Jeremy Darling \& Riccardo Giovanelli}
\affil{Department of Astronomy, Cornell University}
\authoraddr{Department of Astronomy, Cornell University, Ithaca,  NY  14853;
	darling@astrosun.tn.cornell.edu; riccardo@astrosun.tn.cornell.edu}

\begin{abstract}
We present the preliminary results of a survey for OH megamasers
underway at the Arecibo Observatory \footnote{The Arecibo
Observatory is part of the National Astronomy and Ionosphere Center, which 
is operated by Cornell University under a cooperative agreement with the
National Science Foundation.}.  The goals of the survey
are to calibrate the luminosity function of OH megamasers to the 
low-redshift galaxy merger rate ($0.1 < z < 0.2$), and to use the 
enhanced sample of OH megamasers provided by the survey to study 
OH megamaser environments, engines, lifetimes, and structure.  
The survey should double the known OH megamaser sample to roughly 100 
objects.  Survey results will be 
presented in installments to facilitate community access to the data.
Here we report the discovery of 11 OH megamasers and one OH absorber, 
and include upper limits on the isotropic 1667 MHz OH line luminosity of 53 
other luminous infrared
galaxies at $z > 0.1$.  The new megamasers show a wide range of spectral
properties, but are consistent with the extant set of 55 
previously reported objects, only 8 of which have $z > 0.1$.  

\end{abstract}

\section{Introduction}
All known OH megamasers (OHMs) have been observed in
luminous infrared galaxies, strongly favoring the most FIR-luminous, 
the ultraluminous infrared galaxies (ULIRGs) (\cite{baa91}).  
Photometric surveys have shown the ULIRGs to be 
nearly exclusively the product of galaxy mergers (\cite{cle96}).  
VLBI measurements have shown 
that OHMs are ensembles of many masing regions which originate 
in the nuclear regions of (U)LIRGs within scales 
of a few hundred parsecs or less (\cite{dia99}).  
OHM activity requires: (1) high molecular density, (2) a pump
to invert the hyperfine population of the OH ground state, and (3) a
source of 18 cm continuum emission to stimulate maser emission (\cite{bur90})
The galaxy merger environment can supply all of these requirements:  the merger
interaction concentrates molecular gas in the merger nuclei, 
creates strong FIR dust emission from reprocessed starburst
light and AGN activity, and produces radio continuum emission from AGN
or starbursts.  Either the FIR radiation field or
collisional shocks in the molecular gas can invert the OH population
via the pumping lines at 35 and 53 $\mu$m.  Masing can then be stimulated 
by 18 cm continuum emission from starbursts or AGN, or even by spontaneous 
emission from the masing cloud itself (\cite{hen87}).

The FIR luminosity of ULIRGs seems to be correlated with the 
merger sequence phase, based on optical morphology and surface brightness 
profiles (Sanders {\it et al.} 1999).  OHM fraction in ULIRGs is a strong 
function of 
\LFIR (see \cite{baa91} and Figure \ref{success}), which would indicate a 
preferred time during the merger sequence for OHM formation.  
This makes some physical sense, based on the high molecular gas 
density required to produce OHMs ($n_{H_2}=10^{4-7}$ cm$^{-3}$; 
\cite{baa91}).  Early in the merger sequence, infall and concentration
of molecular gas in the nuclear regions is just beginning, whereas late
in the merger sequence, clouds are disrupted by ionizing radiation, 
a superwind phase, or a QSO eruption.  
If OHMs mark a specific phase in major mergers, then they provide
useful tracers of the galaxy merger rate as a function of redshift, 
particularly since they may be observed at cosmological distances with
current instrumentation (\cite{baa89}; \cite{bur90}; \cite{briggs98})

Many studies of the high redshift galaxy luminosity function 
indicate that the galaxy population has undergone either luminosity
evolution, number density evolution, or both since $z\sim0.7$ 
(\cite{lil95}; \cite{ell96}).  
The only way to disentangle the two effects in galaxy evolution is to
directly study the evolution of the galaxy number density, which 
depends on the galaxy merger rate.  Morphological evolution of field 
galaxies hints that mergers may be important:  HST surveys
indicate a morphological evolution of field galaxies, with a larger 
proportion of faint irregular galaxies in the past compared to the 
present (\cite{bri98}).  The connection between gross 
morphological evolution and mergers seems to be supported by surveys
aimed at the detection of the morphological signatures of merger activity.  
Several morphological merger surveys have measured 
an increase in merger fraction with redshift (\cite{lef99}; \cite{pat97}).  
Evolution of the number density of galaxies is a key component of 
hierarchical models of galaxy formation, and measurements of the 
merger rate as a function of redshift should provide meaningful 
constraints on galaxy formation models (see \cite{abr98} for a review).

The merger rate, or merger fraction, of galaxies can be measured 
by surveys targeting the observable signatures of
merger activity, which include:  (1) strong FIR dust emission, 
(2) enhanced star
formation and/or AGN activity, (3) morphological irregularities, such 
as tidal tails, rings, filaments, or shells, and (4) enhanced molecular
gas emission, including OH masing.  

Detection of enhanced FIR dust emission
from ULIRGs is currently limited to $z\lesssim 0.3$ by the sensitivity 
of the {\it Infrared Astronomical Satellite} ({\it IRAS}) at 60 $\mu$m
(Clements {\it et al.}  1999).  Hence, the merger rate of galaxies can be 
measured using ULIRG surveys, but only at very low redshifts.  
High-redshift counterparts to ULIRGs appear to be the sub-mm galaxies 
detected at redshifts of roughly 1--3 (\cite{cow99}; \cite{sma99}).  
Only a few of the sub-mm
galaxies have measured redshifts, a task made difficult by high
optical/UV extinction in these sources.
A meaningful measure of the merger rate based on sub-mm detections 
may be premature at this time. 

The most 
promising method for measuring the galaxy merger rate up to $z=1$ 
relies on the morphological disturbances present in mergers.  Several
groups have used morphological and close pair surveys to measure the 
merger fraction (see \cite{lef99}; \cite{pat97}), and Le F\'{e}vre 
{\it et al.} has
determined a power-law dependence of merger fraction on redshift with 
slope $(1+z)^{3.4}$ up to $z=0.91$.  The approach 
requires high angular resolution and 
high sensitivity to identify the faint debris associated with mergers.  
Flux-limited morphology surveys suffer from the bias produced by the 
brightening associated with galaxy interactions, and the extinction 
produced in advanced mergers as molecular gas and dust is 
concentrated into the central regions of merging galaxies.  Morphological 
surveys are also likely to omit advanced mergers which have a single 
envelope and two closely separated nuclei.  

Finally, one might measure the
merger rate of galaxies using OHMs as tracers of merger activity 
(\cite{bur90}; \cite{briggs98}).  This
is an avenue as yet unexplored, and the role of OHMs in mergers remains
poorly understood.  If one can calibrate the OHM fraction in ULIRGs as a 
function of \LFIR at low redshifts (which can be related to the 
low-redshift merger fraction), then surveys for OHMs at higher redshifts
can yield a measure of the merger rate of galaxies as a function of cosmic time. 
This technique relies on the assumption that OHMs are a constant fraction of 
ULIRGs as a function of cosmic time, which seems to be a reasonable assumption, 
given the weak dependence of metallicity on redshift in the nuclear
regions of massive spiral galaxies.  Using OHMs to trace
merger activity has several advantages over other techniques: 
(1) OHM emission lines 
are detectable at much greater distances than FIR emission with current 
instrumentation, (2) detection of OHMs does not rely on high angular 
resolution, and does not exclude advanced mergers, and (3) OHM detection
favors the large column densities produced in merger environments which
tend to cause extinction in the optical and NIR regime.  

The primary goal of this OHM survey is to determine the incidence of OHMs in
ULIRGs as a function of FIR luminosity, in order to relate the
luminosity function of OHMs to the low redshift galaxy merger rate
($ 0.1< z < 0.2$).  This will allow subsequent workers to estimate 
the galaxy merger rate at higher redshifts from blind or targeted OHM surveys.
Note that previous OHM surveys have obtained estimates of the OHM fraction
in LIRGs, but all are subject to large uncertainties due to small numbers of 
detections (\cite{nor89}; \cite{sta92}; \cite{baa92}; and others).
One might expect the galaxy merger rate 
to roughly follow the evolution of quasars (see \cite{kim98} and \cite{lef99}),
but this notion has not yet been confirmed.  
A useful side benefit of the survey will be a doubling of the OHM sample
and a sixfold increase in the $z>0.1$ OHM sample. 
We intend to use the enhanced sample to study OHM environments,
lifetimes, engines, and structure.  This paper presents the results 
of the preliminary phase of the survey, which targets a FIR selected,
flux-limited sample of about 300 sources, 69 of which have been observed
to date.  Results will be 
presented in installments to facilitate community access to the data.

% Previous searches for OHMs generally do not 
% include lists of all candidates observed or upper limits on OH 
% flux from non-detections, leading subsequent workers to invest
% observing time on repeat observations (there are a few notable exceptions:
%\cite{sta87}, \cite{nor89}, \& \cite{sta92}).  
In this paper, we describe 
the methods used for a new search for OHMs (\S \ref{searchmethods});
we present and discuss the OH detections as well as the non-detections, 
including upper limits on OH luminosity in non-detections (\S \ref{results});
we evaluate the detection rate of our candidate selection methods,
and discuss the new sample of OHMs (\S \ref{discussion}); and we
predict the final outcome of the survey, 
and discuss future prospects (\S \ref{conclusion}).

This paper parameterizes the Hubble constant as $H_\circ = 75\ h_{75}$
km s\minusone\ Mpc\minusone, assumes $q_\circ = 0$, and 
uses $D_L = (v_{CMB} / H_\circ)(1+0.5z_{CMB})$ to 
compute luminosity distances from $v_{CMB}$, the cosmic microwave
background (CMB) rest-frame velocity.  
% Note that the choice $q_\circ = 0$
% biases $D_L$, and hence computed luminosities, low.  At $z=0.2$, luminosity
% distances are $10\%$ lower than in a $q_\circ = 0.5$ cosmology, and 
% luminosities
% are roughly $20\%$ lower.  Care should be taken when comparing FIR or OH line
% luminosities between this paper and previous OHM work.

\section{Search Criteria \& Observations}\label{searchmethods}

Prior to this work, 55 OHMs were reported, which exhibit a wide range of 
characteristics (see \cite{baa98} for a listing of 52 OHMs).  This 
sample revealed a relationship between the FIR luminosity and the 
isotropic OH luminosity
of OHMs which was originally thought to be quadratic ($L_{OH} \propto
L_{FIR}^2$; see \cite{baa89}).  As discussed by \cite{kan96}, 
Malmquist bias affects both OH detection and FIR selection of OHM 
candidates from 
the {\it IRAS} data, and the two datasets are biased differently.  
Hence, the correlation of \LFIR and \LOH with $D_L$ would necessarily
create a correlation between the two variables.  This third 
variable correlation 
must be removed to determine the true correlation between \LFIR and \LOH
for fixed $D_L$.  Kandalian uses the partial correlation
coefficient method for 49 OHMs,
which reveals a $L_{OH}$-\LFIR relationship more linear than quadratic:  
$\log L_{OH} = (1.38 \pm 0.14) \log L_{FIR} - (14.02 \pm 1.66)$.
\cite{dia99} suggest that this relationship represents an admixture
of unsaturated maser emission (quadratic relationship) and saturated
maser emission (linear relationship) from diffuse and compact
OH masing regions, respectively.  In either case, a useful selection 
criterion for locating new OHMs is the requirement for strong FIR emission
from a candidate host, typically $\log (L_{FIR}/L_\odot) \geq 11.0$.

OHM candidates were selected from the Point Source Catalog redshift survey
(PSCz; W. Saunders 1999, private communication), 
supplemented by the NASA/IPAC Extragalactic 
Database\footnote{The NASA/IPAC Extragalactic Database (NED) is operated 
by the Jet Propulsion 
Laboratory, California Institute of Technology, under contract with the 
National Aeronautics and Space Administration.}.  
The PSCz catalog is a flux-limited ({\it IRAS} $f_{60\mu m} > 0.6$ Jy)
redshift survey of 15,000 {\it IRAS} galaxies over $84\%$ of the sky
(see \cite{sau99b}).
We select {\it IRAS} sources which are in the Arecibo sky ($0^\circ <
\delta <
37^\circ$), were detected at 60 $\mu$m, and have $0.1\leq  z \leq 0.45$.  
The lower redshift bound is set to avoid local radio frequency interference
(RFI), while the upper bound is set by the bandpass of the wide L-band receiver
at Arecibo.  No constraints are placed on FIR colors or luminosity.  The 
redshift requirement limits the number of candidates in the Arecibo sky 
to 377.  Of these, 296 are found in the PSCz survey.  The condition that
candidates have $z>0.1$ automatically selects (U)LIRGs if they were detected
by {\it IRAS} at 60 $\mu$m.  The strong influence of \LFIR on OHM fraction
in ULIRGs is the primary reason for our high detection
rate compared to previous surveys (see for example \cite{sta92} and 
\cite{baa92}).  

The upgraded Arecibo radio telescope offers new opportunities for the
detection of OHMs, due to its improved sensitivity, frequency agility, and
instantaneous spectral coverage.  Its large collecting area 
makes it ideal for a survey of spectral lines at the upper end of the 
redshift range of the known OHM sample ($0.1 \leq z \leq 0.3$).  
Detection of OH emission lines
is generally possible in a 4-minute integration, even for $z \simeq 0.2$.  
In roughly 200 hours, we expect to observe about 300 OHM candidates and 
double the sample of OHMs (see \S \ref{conclusion}).  

Observations were performed at Arecibo by nodding on- and off-source in 
4-minute intervals, followed by firing a noise diode after each off-source 
integration.  Spectra were recorded every 6 seconds to facilitate 
RFI removal.  Data was recorded with 9-level sampling in 2 polarizations 
of 1024 channels each, spanning 25 MHz.
The bandpass was centered on 1666.38 MHz (the mean of the 1667 and 1665 
MHz lines), redshifted appropriately for each source.  Strong, sharp RFI 
features
were flagged in the 6-second records using a broad hanning filter and 
a power threshold in the following manner:  (1) each spectrum
was hanning smoothed with 9-point sampling and subtracted from the original 
unsmoothed spectrum, and (2) channels with $> 1\%$ of the total power were
flagged as RFI.  Each 4-minute on-off pair was converted to system 
temperature units, then to flux density units (mJy) 
using a noise diode, which was calibrated to VLA flux standards.  
Baselines were fit and subtracted from each on-off pair and multiple 
pairs were combined via a noise-weighted average.  Polarizations 
were averaged, and the final spectra were hanning smoothed.  The 
frequency resolution after hanning smoothing is 49 kHz.  The uncertainty
in the absolute flux scale is 8$\%$.  

At the time of observations, the L-band wide receiver beam was slightly 
elliptical, with an average HPBW of 3.4$\arcmin$ at 1420.4 $\pm$ 6.25 MHz
(\cite{hei99}\footnote{See the Arecibo Observatory Technical and Operations Memo
99-02 by C. Heiles at http://naic.edu/$\%$7Edonna/performance.htm}).  
At 1500 MHz, the beam would have HPBW of 3.2$\arcmin$.  
The system temperature was 36 K at a zenith angle of 4$^\circ$, and 
increased monotonically to 40 K at ZA$= 18^\circ$.  The gain was 
roughly constant at 9.9 K Jy\minusone\ up to ZA$= 16^\circ$, and decreased
to 9.6 K Jy\minusone\ at ZA$=18^\circ$.  

\section{Results}\label{results}

Observations were made of 68 candidate OHMs with 
$0^\circ < \delta < 37^\circ$, $0.10 \leq z \leq 0.45$, and
$2^h < \alpha < 23^h$ from a set of 235 targets which satisfy these 
selection criteria.  The completion is highest ($68\%$)
in $18^h \leq \alpha \leq 22^h$.  Note
that {\it IRAS} 19084+3719 falls outside of $0^\circ < \delta < 37^\circ$,
but it is included in the lists of non-detections.  

Of the 68 candidates surveyed, 11 new OHMs and 1 new OH absorber were detected.  
Strong upper limits 
for OH luminosity can be placed on 53 non-detections.  Three candidates 
were rejected due to RFI ({\it IRAS} F02054+0835,
F13380+3339, and F15438+0438) and {\it IRAS} F15599+0206 
was set aside due to strong radio continuum (see discussion 
of this selection effect below).  

\subsection{Non-Detections}

Tables \ref{nondetectFIR} and \ref{nondetectOH} list respectively 
the optical/FIR and radio properties of the 54
OH non-detection ULIRGs.  We can predict the expected \LOH for the 
OHM candidates based on the 
$L_{OH}$-\LFIR relation computed by Kandalian (1996; see \S 
\ref{searchmethods}) and compare this figure to upper limits on \LOH derived
from observations for a rough measure of the confidence of the 
non-detections.  Table \ref{nondetectFIR} lists the optical redshifts
and FIR properties of the non-detections in the following format.
Column (1):  {\it IRAS} Faint Source Catalog (FSC) name.  
Columns (2) and (3):  Source coordinates (epoch B1950.0) 
from the FSC, or the Point Source Catalog (PSC) if not available in the FSC.  
Columns (4), (5) and (6):  Heliocentric optical redshift, 
reference, and corresponding
velocity.  Uncertainties in velocities are listed whenever they are 
available.  
Column (7):  Cosmic microwave background rest-frame velocity.  This is
computed from the heliocentric velocity using the solar motion with respect
to the CMB measured by \cite{lin96}:  $v_\odot = 368.7 \pm 2.5$ km s\minusone\
towards $(l,b) = (264\fdg31 \pm 0\fdg16 , 48\fdg05 \pm 0\fdg09)$.
Column (8):  Luminosity distance computed from $v_{CMB}$ via 
$D_L = (v_{CMB} / H_\circ)(1+0.5z_{CMB})$, assuming $q_\circ = 0$.  
Columns  (9) and (10):  {\it IRAS} 60 and 100 $\mu$m flux 
densities in Jy.  FSC flux densities are listed whenever they are 
available.  Otherwise, PSC flux densities
are used.  Uncertainties refer to the last digits of each measure, and upper 
limits on 100 $\mu$m flux densities are indicated by a ``less-than'' symbol. 
Column (11):  The logarithm of the far-infrared luminosity in units
of $L_\odot$.  \LFIR is computed following the prescription of \cite{ful89}:  
\LFIR$ = 3.96\times 10^5 D_L^2 (2.58 f_{60} + f_{100})$, 
where $f_{60}$ and $f_{100}$ are the 60 and 100 
$\mu$m flux densities expressed in Jy, 
$D_L$ is in Mpc, and \LFIR is in units of $L_\odot$.  
If $f_{100}$ is only available as an upper limit, the permitted range
of \LFIR is listed.  The lower bound on \LFIR is computed for $f_{100}=0$ mJy,
and the upper bound is computed with $f_{100}$ set equal to its upper limit.
The uncertainties in $D_L$ and the flux densities 
typically produce an uncertainty in $\log L_{FIR}$ of $0.01$.  

Table \ref{nondetectOH} lists the 1.4 GHz 
flux density and the limits on OH emission of the non-detections in 
the following format:
Column (1):  {\it IRAS} FSC name, as in Table \ref{nondetectFIR}.
Column (2):  Heliocentric optical redshift, as in Table \ref{nondetectFIR}.
Column (3):  \LFIR, as in Table \ref{nondetectFIR}.
Column (4):  Predicted isotropic OH line luminosity, 
$\log L_{OH}^{pred}$,
based on the Malmquist bias-corrected $L_{OH}$-\LFIR relation 
determined by \cite{kan96} for 49 OHMs:  
$\log L_{OH} = (1.38\pm0.14) \log L_{FIR} - (14.02\pm 1.66)$ 
(see \S \ref{searchmethods}).
Column (5):  Upper limit on the isotropic OH line luminosity, 
$\log L_{OH}^{max}$.  
The upper limits on \LOH are computed from the RMS noise of the non-detection
spectrum assuming a ``boxcar'' line profile of rest frame width 
$\Delta v = 150$ km s\minusone\ and height 1.5 $\sigma$: 
$\log L_{OH}^{max} = \log (4 \pi D_L^2\ 1.5 \sigma [\Delta v/c] 
[\nu_\circ /(1+z)])$.   The assumed rest frame width $\Delta v = 150$ km
s\minusone\ is the average FWHM of the 1667 MHz line of the known OHM sample.
Column (6):  On-source integration time, in minutes.  
Column (7):  RMS noise values in flat regions of the 
non-detection baselines, in mJy, after spectra were hanning smoothed 
to a spectral resolution of 49 kHz.
Column (8):  1.4 GHz continuum fluxes, from the NRAO 
VLA Sky Survey (\cite{con98}).  If no continuum source lies within 
30$\arcsec$ of the {\it IRAS} coordinates, an upper limit of 5.0 mJy 
is listed.  
Column (9):  Optical spectroscopic classification, if available.  
Codes used are:  ``S2'' = Seyfert type 2;  ``S1.9'' = Seyfert type 1.9;
``S1.5'' = Seyfert type 1.5;  
``S1'' = Seyfert type 1;  ``H'' = \ion{H}{2} region (starburst);  
and ``L'' = low-ionization emission region (LINER).  References
for the classifications are listed in parentheses and included at the 
bottom of the Table.
Column (10):  Source notes, listed at the bottom of the Table.

An estimate of the confidence 
of non-detections among the sample can be found from a comparison of 
$L_{OH}^{pred}$ to $L_{OH}^{max}$.  Note, however, 
that the scatter in the $L_{OH}$--$L_{FIR}$ relation is quite large:  roughly 
half an order of magnitude in \LFIR and one order of magnitude in \LOH (see
Kandalian 1996).  Among the non-detections, 6 out of 53 galaxies have 
$L_{OH}^{pred} < L_{OH}^{max}$, indicating that longer integration times 
are needed to confirm these non-detections.
8 out of 53 candidates have $L_{OH}^{max}$ within the range of $L_{OH}^{pred}$
set by an upper limit on $f_{100}$ (we exclude two sources with strong
standing wave patterns from
these tallies because their RMS does not reflect true noise levels).  
Integration times were a compromise between efficient use of telescope
time and the requirement for a meaningful upper limit on \LOH of
non-detections.
Given the scatter of identified OHMs about the $L_{OH}$--$L_{FIR}$
relation, we estimate that there are perhaps 4 additional OHMs
among the non-detections, but this estimate relies on uncertain
statistics of small numbers.  A thorough analysis of the detection
completeness will be performed once the survey is complete.  

Strong continuum sources ($S_{1.4 GHz} \gtrsim 100$ mJy)
produce standing waves between the Gregorian dome and the primary 
reflector, which frustrates detection of 1--10 mJy spectral lines.
It may be possible to remove standing waves through a ``double-switching''
process by observing a pure continuum source of comparable flux density.
For the present, the source which produced strong standing waves was
set aside for later re-observation, and no upper limit on \LOH was
calculated.
The correlation between radio continuum and FIR flux density of megamaser 
hosts implies a correlation between radio continuum and OH 
megamaser emission due to the $L_{OH}$-\LFIR relation (see \cite{sta92}), 
indicating that the removal of strong continuum sources from the 
sample might create a strong selection effect.  
Of the 4 observed OHM candidates exhibiting standing waves, 
however, only two have
$S_{1.4GHz} > 100$ mJy, as determined from the NRAO VLA Sky Survey 
(\cite{con98}).  The other two are contaminated by continuum sources
close to the target or falling in a side lobe of the main beam.  These two 
were not excluded by their own properties, but by a random process.  Hence,
their exclusion should not bias the survey.

\subsection{OH Detections}
Tables \ref{detectFIR} and \ref{detectOH} list respectively the 
optical/FIR and radio properties of the 12 new OH detections.  
Spectra of the 11 OHMs and the single OH absorber appear
in Figure \ref{spectra}.  
The column headings of Table \ref{detectFIR} are identical to those of Table
\ref{nondetectFIR}.  Table \ref{detectOH} lists the OH emission/absorption
properties and 1.4 GHz flux density of the OH detections in the format.
Column (1):  {\it IRAS} FSC name.
Column (2):  Measured heliocentric velocity of the 1667.358 MHz 
line, defined by the center of the FWHM of the line.  The uncertainty in the
velocity of the line center is estimated assuming an uncertainty of $\pm 1$
channel ($\pm 49$ kHz) on each side of the line.  
Column (3):  On-source integration time in minutes.  
Column (4):  Peak flux density of the 1667.359 MHz OH line in mJy.
Column (5):  Equivalent width-like measure in MHz.  
$W_{1667}$ is the ratio of the integrated 1667.359 MHz line flux to its 
peak flux.  
Column (6):  Observed FWHM of the 1667.359 MHz OH line in MHz.
Column (7):  Rest frame FWHM of the 1667.359 MHz OH line in km 
s\minusone.
The rest frame width was calculated from the observed width using the relation
$\Delta v_{rest} = c (1+z) (\Delta \nu_{obs} / \nu_\circ)$.
Column (8):  Hyperfine ratio, defined by $R_H = F_{1667}/F_{1665}$, where
$F_\nu$ is the integrated flux density across the emission/absorption line centered on 
$\nu$.  $R_H = 1.8$ in thermodynamic equilibrium, and increases as the degree
of saturation of masing regions increases.  In many cases, the 1665 MHz
OH line is not apparent, or is blended into the 1667 MHz OH line, and a good
measure of $R_H$ becomes difficult without a model for the line profile.  It is
also not clear that the two lines should have similar profiles, particularly if the
lines are aggregates of many emission regions in different saturation states. 
Some spectra allow a lower limit to be placed on $R_H$, indicated by a greater than
symbol.  Blended or noisy lines have uncertain values of $R_H$, and are indicated 
by a tilde.  For {\it IRAS} 16300+1558, RFI makes any estimate of the hyperfine 
ratio impossible.  
Column (9):  Logarithm of the FIR luminosity, as in Table \ref{detectFIR}.
Column (10):  Predicted OH luminosity, $\log L_{OH}^{pred}$, as in Table 
\ref{nondetectOH}.
Column (11):  Logarithm of the measured isotropic OH line luminosity, 
which includes the 
integrated flux density of both the 1667.359 and the 1665.4018 MHz lines.
Note that $L_{OH}^{pred}$ is generally 
less than the actual \LOH detected (9 out of 11 detections), 
as expected from Malmquist bias.  
Column (12):  1.4 GHz continuum fluxes, from the NRAO VLA Sky Survey 
(\cite{con98}).  If no continuum source lies within 30$\arcsec$ of the {\it IRAS} 
coordinates, an upper limit of 5.0 mJy is listed.  

The spectra of the OH detections are shown in Figure \ref{spectra}.  
The abscissae and inset redshifts refer to the optical heliocentric 
velocity, and the arrows indicate the expected velocity of the 1667.359 
({\it left}) and 1665.4018 ({\it right}) MHz lines based on the optical redshift, 
with error bars indicating
the uncertainty in the redshift.  The spectra refer to  1667.359 MHz as 
the rest
frequency for the velocity scale.  {\it IRAS} 19154+2704 has no velocity 
uncertainty available in the literature.  

In order to quantitatively identify somewhat dubious 1665 MHz OH line
detections, we compute the autocorrelation function (ACF) of each spectrum and 
locate the secondary peak (the primary peak corresponds to zero offset, or
perfect correlation).   Any correspondence of features between the two main 
OH lines will enhance the second autocorrelation peak and allow us to 
unambiguously identify 1665 MHz lines based not strictly on spectral 
location and peak flux, but on line shape as well.
The secondary peak in the ACF of each spectrum, when present, 
is indicated by a small solid line.  We expect 
the offset of the secondary peak to be equal to the separation
of the two main OH lines, properly redshifted:  (1.9572 MHz)$/(1+z)$.  The 
{\it expected} location of the secondary ACF peak is indicated
in Figure \ref{spectra} by a small dashed line over each spectrum.  
Both the expected and actual secondary peak correlation positions are plotted 
offset with respect to the center of the 1667 MHz line, as defined by the 
center of the FWHM, rather than the peak flux.

The expected relationship between the hyperfine ratio and the value of the 
secondary peak in the ACF should provide an upper limit on the actual size
of the secondary peak.  In the limit of perfect correspondence of features
between between the two main OH lines, the ACF second peak value becomes
$R_H/(1+R_H^2)$.  Hence, one might measure the correspondence between the 
1667 and 1665 MHz lines by normalizing
the actual secondary peak value to this upper limit.  We will explore the
utility of this line correspondence measure when we obtain a larger sample
of OHMs.

We make some observations and measurements specific to individual OH detections
as follows.

\noindent{\bf 06487+2208:} This merger was originally misclassified as 
Galactic cirrus by \cite{str92},
but a redshift was measured in the QDOT survey (\cite{lu95}).  
The OH spectrum shows both the 1667 and 1665 MHz lines, and a 
strong correspondence of features between the two.  The ACF has a 
strong secondary peak and is in excellent agreement with the predicted 
location of the 1665 MHz line based on both the optical redshift and
the frequency of the 1667 MHz line (the dotted and solid lines
in Figure \ref{spectra} overlap).  
An HST WFPC2 archive image\footnote{Based on observations 
made with the NASA/ESA Hubble Space Telescope, obtained from the data 
archive at the 
Space Telescope Science Institute. STScI is operated by the Association of 
Universities 
for Research in Astronomy, Inc. under NASA contract NAS 5-26555.} shows a 
disturbed morphology and multiple nuclei (see Figure \ref{hst}).  
A Palomar 200'' telescope spectrum indicates that the bright nuclei of
this source show a composite of \ion{H}{2} and LINER characteristics, 
based on the Osterbrock spectral line ratio classification method 
(\cite{ost89}; \cite{vei95}).  The spectral line data and analysis will
be presented in a later paper when we have obtained optical spectra of a 
large number of OHMs.  
The nuclei of this source may have different properties, 
and result in a net blend of emission region types in a 
spectrum which is spatially unresolved from the ground.  

\noindent {\bf 16300+1558:} This is the second most distant OHM known, at 
$z_\odot=0.2417$ ({\it IRAS} 14070+0525 has $z_\odot=0.2644$ --- see 
\cite{baa92b}).  Optically, Kim, Veilleux, \& Sanders (1998) 
classify it as a LINER.  The observed bandpass is disturbed by several 
RFI features, including a feature which obliterates the 1665 MHz line 
region.  Hence, no
hyperfine ratio can be measured for this source, and no autocorrelation 
analysis can be performed.  The 1667 MHz line velocity agrees with 
the optical redshift to within the uncertainty of 64 km s\minusone.
Note that the shape of the 1667 MHz line is strikingly similar 
to that of {\it IRAS} 06487+2208.  Both have a double-peaked structure, which 
suggests either multiple maser sites or a rotating masing torus, as in 
III Zw 35 (\cite{dia99}).  {\it IRAS} 16300+1558
is the most FIR-luminous source observed in our candidate list to date, but
its OH luminosity falls short of the \LOH predicted from $L_{FIR}$.  
Re-observation in quieter RFI conditions would be desirable in order
to detect the 1665 MHz line and confirm the OH luminosity of this exceptional
object.

\noindent {\bf 17539+2935:}
The primary OH line detected in this object corresponds with the optical 
redshift prediction for the 1667 MHz line.  No corresponding 1665 MHz 
line is visible in the spectrum.  A secondary OH line of marginal
signal-to-noise is blue-shifted with 
respect to the primary line by
462 km s\minusone\ in the rest frame.  The second OH line suggests a 
double-nucleus host, with each nucleus producing OH emission.
The Digitized Sky Survey\footnote{Based on 
photographic data obtained using Oschin Schmidt Telescope on Palomar 
Mountain. The Palomar Observatory Sky Survey was funded by the National 
Geographic Society. The Oschin Schmidt Telescope is operated by the 
California Institute of Technology and Palomar Observatory. The plates 
were processed into the present compressed digital format with their
permission. The Digitized Sky Survey was produced at the Space Telescope 
Science Institute (STScI) under U.S. Government grant NAG W-2166.} 
image reveals an extended spidery morphology 
which could be attributed to expelled merger debris.
More integration time on this source is needed to detect the 1665 MHz 
line and to confirm the blue-shifted component.  The hyperfine ratio 
can only be given an upper limit for this 
source.  We assume a square profile of width equal to the 1667 MHz line
width and height 1$\sigma$ to obtain $R_H \geq 2.9$.

\noindent {\bf 18368+3549:} This source shows a broad OH line profile 
($\Delta v_{1667} = 421$ km s\minusone) which includes several sharp 
components.  The hyperfine ratio listed in Table \ref{detectOH} 
assumes that the small ``shoulder'' on the high velocity side of the
main line is the 1665 MHz line.  There is likely some blending
of lines, which would tend to decrease $R_H$.  
The Digitized Sky Survey image shows an extended source 
with irregular morphology, but 
IR and optical imaging by \cite{mur96} shows this object to have 
a single nucleus, or if double, a nuclear separation of $<0.8\arcsec$ 
($<1.5$ kpc) in K band (the upper limit is set by the seeing).  
The r-band image 
shows a slightly disturbed morphology, while the K-band image is nearly 
unresolved.  
As expected of any OHM host, {\it IRAS} 18368+3549 is rich in molecular gas
which is concentrated in a small nuclear volume.
\cite{sol97} measure a CO(1--0) line width of 330 km 
s\minusone, estimate a H$_2$ mass of 3.9$\times 10^{10} M_\odot$, 
and derive a blackbody radius of 299 pc.  

\noindent {\bf 18588+3517:} This OHM has no broad emission 
component evident, and appears to be dominated by a pair of sharp features
at 1667 MHz.  The 1667 and 1665 MHz lines show a remarkable correspondence  
which indicates a much larger hyperfine ratio for the 
main peak than for the smaller, lower velocity peak.  Hence,
the main peak is the more saturated of the two.  The optical redshift
is significantly larger ($\sim 300$ km s\minusone) than the observed line
velocity.  This could be due to an underestimated error in the optical 
redshift, or a double nucleus source:  the OH emission may come from an 
obscured nucleus, while the optically brighter nucleus may be responsible for
the observed optical redshift.  This explanation seems plausible because OH 
maser emission favors high gas column densities, but optical line observations 
omit high extinction environments.  

\noindent {\bf 19154+2704:} This is an OH absorber, with $R_H = 1.81$, which 
is consistent with the hyperfine ratio in thermodynamic equilibrium 
($R_H = 1.8$).  There is good agreement between the optical redshift and
the observed velocity of the absorption lines (note that the optical
redshift has no uncertainty available in the literature).  The 1665 MHz 
absorption line is strong, and has a velocity in good agreement with 
the ACF secondary peak and line separation predictions.  The host has a
fairly strong 1.4 GHz continuum (64 mJy), which creates some mild standing
waves in the baseline.  

\noindent {\bf 20248+1734:} The OH spectrum of this source shows significant 
continuum emission within the 3.2$\arcmin$ beam
($\sim 75$ mJy at 1.487 GHz).  The closest NVSS source to 
{\it IRAS} 20248+1734 is 1.9
arcminutes distant and has a 1.4 GHz flux of 209 mJy (\cite{con98}).  
Contamination from the NVSS source may be responsible for 
the mild standing waves evident in the OH spectrum.  These standing waves 
contaminate
the emission lines and may be creating an artificial 1665 MHz line, as 
well as broadening the 1667 MHz line.  A higher signal-to-noise spectrum 
is required to remove the contamination.
The ACF and 1667 MHz line centroid both identify a spectral feature as
the 1665 MHz line, but this feature is not significantly different
from other baseline standing waves.  The optical redshift differs
significantly from the observed OH line velocity.  

\noindent {\bf 20286+1846:} This OHM shows both broad and sharp spectral
components.  The very broad ``shoulders'' of the OH emission 
(rest frame $\Delta v = 1040$ km s\minusone\ at $10\%$ peak flux) 
suggest a spatially 
extended region of diffuse emission.  The total OH luminosity emitted by 
this object is unusually strong:   $\log (L_{OH}/L_\odot) = 3.33$, more than 
one order of magnitude 
larger than predicted from $L_{FIR}$: $\log (L_{OH}^{pred}/L_\odot) = 2.15$.  
Identification of the 1665 MHz line is difficult, due to blending of
the lines.  Moving from high velocity to low, the first and second peaks
have the correct separation from the third and fourth peaks, respectively, 
to be 1665 MHz lines.  The optical redshift and 1667 MHz line centroid both
agree with this identification.  Assuming this to be true, we measure
a hyperfine ratio of $R_H=4.4$.  On the other hand, the low-velocity
shoulder indicates that it is possible to have broad 1667 MHz emission
far from the peak, so all of the high-velocity emission could be strictly
1667 MHz.  Hence, we say $R_H \geq 4.4$.  Since OH masing is beamed 
emission which relies on a pump and stimulant, it is possible that only 
blue-shifted
foreground emission should appear along the line of sight.  If, however,
there is a torus of emission, high and low velocity features might be
expected to be symmetrical about the center line.  There is currently 
not enough information to break the degeneracy in the physical emission 
configurations of this OHM.

\noindent {\bf 20450+2140:} No broad component is evident in the OH 
spectrum of this source, nor
is any 1665 MHz emission detected.  The ACF and expected line separation
both indicate 
a marginal feature to be the 1665 MHz line.  This feature is identical
to several other noise features in the baseline, so we can only compute
a lower limit on the hyperfine ratio:  $R_H \geq 6.2$.
The spectrum suggests two sharp emission components, but could in fact be
a single line modified by detector noise.

\noindent {\bf 21077+3358:} The OH spectrum shows both broad and sharp 
components, including broad ``shoulders'' similar to {\it IRAS} 
20286+1846 (rest frame $\Delta v = 1330$ km s\minusone\ at $10\%$ peak flux).  
The high velocity shoulder may either be broad 1667 MHz
emission or the 1665 MHz line.  If we assume the latter, then 
$R_H = 7.4$.  If there is any 1667 mixing into this component, which 
seems probable given the appearance of the low 
velocity shoulder, then $R_H$ will be larger.  Hence, 
we conclude that $R_H \geq 7.4$.  The ACF makes no prediction for
the location of the 1665 MHz line (it shows no secondary peak), but 
the expected line separation indicates a 1665 MHz line centroid 
position in the center of the high-velocity shoulder.
Poor subtraction of Galactic HI produced the feature at 52000 km s\minusone.  

\noindent {\bf 21272+2514:}  This OHM is the third most luminous known 
($\log (L_{OH}^{pred}/L_\odot) = 3.63$) at a redshift of $z=0.1508$.  
Its observed \LOH is more than
one order of magnitude greater than we predict from $L_{FIR}$.  
The spectrum exhibits multiple
peaks and a rest frame width of 849 km s\minusone\ at 10$\%$ of peak flux.
The 1665 MHz line can be identified based on
the optical redshift.  The measurement of $R_H = 13.7$ given in 
Table \ref{detectOH} assumes that there are no 1665 MHz lines associated
with the 1667 MHz peaks with velocities below the optical redshift.  
As indicated in Figure \ref{spectra}, the centroid of the 1667 MHz 
emission falls below the optical redshift, and the ACF shows no secondary
maximum.  

\noindent {\bf 22116+0437:}  The Digitized Sky Survey
image of the host galaxy shows a possible galaxy pair in a single envelope.
The two spectral lines show good correspondence with the optical redshift
predictions.  Despite the low luminosity of this OHM, \LOH is 
greater than $L_{OH}^{pred}$ by a factor of $\sim$2.  This is the third
most distant OHM detected to date, at $z=0.1939$.

\section{Discussion}\label{discussion}

Detection of 11 new OHMs out of the 65 candidates observed yields a 
success rate of
1 in 6 ($17\%$).  Figure \ref{success} plots \LFIR versus FIR color
($f_{100\mu m} / f_{60\mu m}$) for all 65 candidates, plus one previously
known OHM included in the survey sample which was 
observed for a system check at the telescope.  As indicated by the histograms
in the right-hand panels, there is a strong tendency for OHMs to appear in the
most FIR-luminous LIRGs.  Any explicit lower bound on \LFIR would 
increase the detection rate (we have no such \LFIR selection criterion), 
but we hesitate to impose a 
constraint which depends on choice of cosmology.  Also, we have low 
confidence in the non-detections in the lowest \LFIR bin because 
they generally have $L_{OH}^{max} > L_{OH}^{pred}$.  There may yet be 
OHMs lurking among these non-detections.
A FIR color selection criterion would also boost
the detection rate, but would cause the loss of some detections.  
{\it Any} color selection criterion would exclude the 63
{\it IRAS} galaxies in the survey sample which are undetected at 
100 $\mu$m.  Many of these have strong 60 $\mu$m fluxes, and are thus 
interesting survey targets.  

The new OHMs show a wide range of properties.  They span two decades in 
OH luminosity, and range from very narrow (64 km s\minusone) single lines
to broad (421 km s\minusone) complicated multi-line ensembles, to sharp
lines atop broad bases of emission.  The sample shows diverse hyperfine
ratios, with hints that the ratio may vary within a single source, indicating
a range of maser saturation states.  None of the new OHMs are strong 
radio continuum sources (the strongest is 21 mJy), and there does not 
appear to be a correlation between OH flux and radio continuum 
flux.  There appears to be a positive relationship between the OH 
luminosity of the
new OHMs and the 1667 MHz line widths, opposite the trend
found by Staveley-Smith {\it et al.} (1992) and \cite{kan96}.  This relationship will be explored
further with a larger sample.  Only two of the new
OHMs have optical spectral classifications (\cite{kim98b} classify 
{\it IRAS} 16300+1558 as a LINER, and we classify {\it IRAS} 06487+2208
as a LINER/\ion{H}{2} mixture), but
the non-detections cover the full range of spectral types, from \ion{H}{2} regions
to LINERs to all Seyfert classifications.  Baan {\it et al.} (1998) 
classified the
bulk of the known sample of OHMs, and found a tendency for OHMs to occur
in active nuclei.  Optical classification of the new higher redshift
sample is under investigation and will be reported when the study is complete.

\section{Conclusions}\label{conclusion}

We have demonstrated the ability of the upgraded Arecibo 
telescope to detect new OH megamasers with a high success rate (1 in 6) 
in the highest OHM redshift regime in short integration times.  The 
detection rate is due in part to the
improved frequency agility and sensitivity of the upgraded Arecibo 
telescope, and in part due to completion of {\it IRAS} galaxy redshift
surveys (most notably the PSCz).  

The survey will produce a set of about 50 new OHMs in short
order, which will double the sample and greatly enhance the 
population at the highest redshifts.  The new sample will be used
to explore the 
physics of OHM phenomena in a statistically meaningful manner and to 
evaluate theoretical models of OHM environments and excitation mechanisms.

The OHM detection rate of our flux-limited survey sample 
will produce an accurate measure of the incidence of OHMs in ULIRGs as a
function of host properties which is much less subject to the small number
statistics than previous surveys.  Subsequent searches for
OHMs will be able to indirectly measure the FIR luminosity function
of ULIRGs at various redshifts, which can be related to the merger rate
of galaxies as a function of cosmic time.  

\acknowledgements

The authors are very grateful to Will Saunders
for access to the PSCz catalog, to Liese van Zee and Martha
Haynes for providing the optical spectrum of {\it IRAS} 06487+2208,
and to the staff of NAIC for observing assistance and support.  

This research was supported by Space Science Institute archival grant 
8373 and made use of the NASA/IPAC Extragalactic Database (NED) 
which is operated by the Jet Propulsion Laboratory, California
Institute of Technology, under contract with the National Aeronautics 
and Space Administration.  

We acknowledge the use of NASA's SkyView facility\footnote{At 
http://skyview.gsfc.nasa.gov.} located at NASA Goddard Space Flight Center.

\clearpage

\begin{figure}
\caption{New OH megamasers/OH absorber discovered in ULIRGs.  
Abscissae and inset redshifts refer to the optical heliocentric velocity.
Spectra use the 1667.359 MHz line as the rest frequency for the 
velocity scale.  
Arrows indicate the expected velocity of the 1667.359 (left)
and 1665.4018 (right) MHz lines based on the optical redshift, with 
error bars indicating the uncertainty in the redshift.  Solid vertical
lines indicate the location of the secondary maximum in the autocorrelation
function, and dashed vertical lines indicate the expected position of the 
1665 MHz line, based on the centroid of the 1667 MHz line.  The 
dotted baselines indicate the shape (but not the absolute magnitude) 
of the baselines subtracted from the calibrated spectra.
The properties of these megamasers are listed in Tables \ref{detectFIR}
and \ref{detectOH}.\label{spectra}}
\end{figure}

\begin{figure}
\caption{This HST WFPC2 archive image$^4$ of {\it IRAS} 06487+2208 
($z_\odot = 0.1437$) taken 
with the F814W filter ($\lambda = 7940$\AA, $\Delta\lambda 
= 1531$\AA) reveals a disturbed morphology and multiple nuclei.  Total 
integration time was 800 seconds, taken in two exposures of 400 seconds 
for cosmic ray removal.  
Coordinate labels are in epoch J2000.  The disturbed morphology is
not evident in ground-based observations with $\sim 1.5\arcsec$ 
seeing.  \label{hst}}
\end{figure}

\begin{figure}
\caption{Observed OH Megamaser Candidates.  The two left panels show \LFIR
versus FIR color for candidates observed to date, and the two right panels
show the OHM fraction as a function of $L_{FIR}$.  Filled circles mark OHMs, 
empty circles mark non-detections, and the crossed circle marks the OH
absorber.  Points with error bars are non-detections at 100 $\mu$m.  
Vertical error bars indicate the possible range of $L_{FIR}$, constrained
by $f_{60\mu m}$ and an upper limit on $f_{100\mu m}$.  Horizontal arrows
indicate upper limits on FIR color.  Inset percentages indicate the OHM
fraction for each sector delineated by the dashed lines.  The upper panels
plot all 65 candidates observed, plus one known OHM reobserved to check
the observing setup in April 1999.  The lower panels plot the 41 objects
with detected $f_{100\mu m}$.  The inset numbers follow the key:  
N $= $ Observed (OHMs, Non-Detections).  We use $H_\circ = 75$ km s\minusone\
Mpc\minusone.
\label{success}}
\end{figure}

\clearpage

\begin{deluxetable}{cccccrrcrcc} 
\scriptsize
\tablecaption{OH Non-Detections:  Optical Redshifts and FIR Properties\label{nondetectFIR}}
\tablewidth{0pt}
\tablehead{
\colhead{{\it IRAS} Name} &  \colhead{$\alpha$} & \colhead{$\delta$} & \colhead{$z_\odot$}  
& \colhead{Ref}& \colhead{$v_\odot$} & \colhead{$v_{CMB}$} 
& \colhead{$D_L$} & \colhead{$f_{60}$} & \colhead{$f_{100}$} & 
\colhead{$\log L_{FIR}$} \nl
\colhead{FSC} & \colhead{B1950} & \colhead{B1950} &  & \colhead{} & \colhead{km s\minusone} &
\colhead{km s\minusone} &  \colhead{$h^{-1}_{75}$ Mpc} &  \colhead{Jy} &  
\colhead{Jy} & \colhead{$h^{-2}_{75} L_\odot$} \nl
\colhead{(1)}& \colhead{(2)}& \colhead{(3)}& \colhead{(4)}& \colhead{(5)}& 
\colhead{(6)}& \colhead{(7)}& \colhead{(8)}& \colhead{(9)}& \colhead{(10)}& 
\colhead{(11)}
 }
\startdata
02290+3139 & 02 29 05.6 & +31 39 28 & 0.2115 & 2 & 63412(105)
& 63188(110) & 931(2) & 0.567(\phn51) & 1.39(28) & 11.99 \nl
03477+2611 & 03 47 43.3 & +26 11 55 & 0.1494 & 2 & 44779(196)
& 44645(199) & 640(3) & 0.711(\phn50) & 1.36(23) & 11.71 \nl
03533+2606 & 03 53 19.9 & +26 06 14 & 0.1883 & 7 & 56451(\phn\phn\phn)
& 56324(\phn37) & 822(1) & 0.414(\phn58) & $<3.11$ & 11.46--12.05 \nl
04046+1011 & 04 04 44.6 & +10 11 55 & 0.1845 & 17 & 55312(150)
& 55204(154) & 804(2) & 0.475(\phn38) & $<4.38$ & 11.50--12.16 \nl
08559+1053 & 08 55 58.8 & +10 53 02 & 0.1480 & 1 & 44369(\phn70)
& 44661(\phn74) & 640(1) & 1.119(\phn67) & 1.95(25) & 11.89 \nl
13542+1040 & 13 54 18.1 & +10 40 50 & 0.1234 & 3 & 37001(\phn46)
& 37263(\phn52) & 528(1) & 0.797(\phn88) & 0.60(15) & 11.47 \nl
14202+2615 & 14 20 16.0 & +26 15 43 & 0.1590 & 1 & 47667(\phn70)
& 47868(\phn76) & 689(1) & 1.492(104) & 1.99(18) & 12.04 \nl
14203+3005 & 14 20 19.4 & +30 05 58 & 0.1141 & 10 & 34202(\phn26)
& 34393(\phn39) & 485(1) & 0.960(\phn77) & 1.39(19) & 11.56 \nl
15445+3312 & 15 44 32.1 & +33 12 57 & 0.1558 & 11 & 46710(\phn81)
& 46796(\phn87) & 673(1) & 0.383(\phn34) & 0.86(15) & 11.52 \nl
15543+3013 & 15 54 22.7 & +30 13 22 & 0.1213 & 11 & 36360(\phn81)
& 36439(\phn88) & 515(1) & 0.393(\phn39) & 0.68(17) & 11.25 \nl
15597+3133 & 15 59 47.3 & +31 33 27 & 0.1437 & 11 & 43070(\phn81)
& 43140(\phn88) & 617(1) & 0.466(\phn42) & 1.11(16) & 11.54 \nl
16045+2733 & 16 04 33.8 & +27 33 03 & 0.1139 & 10 & 34147(\phn22)
& 34217(\phn40) & 482(1) & 0.828(\phn58) & 1.77(23) & 11.56 \nl
16121+2611 & 16 12 08.7 & +26 11 47 & 0.1310 & 12 & 39273(\phn\phn\phn)
& 39335(\phn34) & 559(1) & 0.191(\phn42) & $<0.41$ & 10.79--11.05 \nl
16156+0146 & 16 15 35.2 & +01 46 42 & 0.1320 & 1 & 39573(\phn70)
& 39657(\phn79) & 564(1) & 1.126(\phn68) & 1.00(22) & 11.69 \nl
16284+2817 & 16 28 29.2 & +28 17 17 & 0.0970 & 10 & 29080(\phn24)
& 29116(\phn42) & 407(1) & 1.109(\phn55) & 0.88(25) & 11.39 \nl
16474+3430 & 16 47 24.2 & +34 30 18 & 0.1115 & 8 & 33418(\phn47)
& 33422(\phn58) & 470(1) & 2.272(114) & 2.88(20) & 11.88 \nl
17030+0457 & 17 03 02.6 & +04 57 45 & 0.1190 & 6 & 35675(300)
& 35681(302) & 504(5) & 0.603(\phn42) & $<1.84$ & 11.19--11.53 \nl
17156+1238 & 17 15 36.9 & +12 38 18 & 0.1130 & 7 & 33876(\phn\phn\phn)
& 33857(\phn36) & 477(1) & 0.778(\phn54) & $<1.26$ & 11.26--11.47 \nl
17490+2659 & 17 49 05.8 & +26 59 46 & 0.1453 & 15 & 43560(\phn\phn\phn)
& 43484(\phn34) & 622(1) & 0.466(\phn42) & $<1.37$ & 11.27--11.60 \nl
17574+0629 & 17 57 26.4 & +06 29 17 & 0.1096 & 8 & 32860(\phn44)
& 32779(\phn57) & 461(1) & 2.075(145) & $<5.48$ & 11.65--11.96 \nl
18030+0705 & 18 03 01.6 & +07 05 39 & 0.1458 & 8 & 43708(\phn56)
& 43618(\phn67) & 624(1) & 0.838(\phn75) & 4.40(48) & 12.00 \nl
18040+2141 & 18 04 06.2 & +21 41 06 & 0.1016 & 13 & 30454(\phn24)
& 30357(\phn42) & 425(1) & 1.474(133) & 2.35(35) & 11.64 \nl
18147+1553 & 18 14 45.2 & +15 53 37 & 0.1024 & 13 & 30695(207)
& 30584(210) & 429(3) & 1.275(115) & 2.04(22) & 11.59 \nl
18222+1440 & 18 22 15.8 & +14 40 13 & 0.1262 & 9 & 37821(\phn\phn\phn)
& 37699(\phn34) & 534(1) & 1.009(\phn91) & $<1.93$ & 11.47--11.71 \nl
18315+2249 & 18 31 30.6 & +22 49 40 & 0.1310 & 2 & 39287(131)
& 39151(135) & 556(2) & 0.868(\phn61) & $<2.06$ & 11.44--11.72 \nl
18585+2148 & 18 58 32.2 & +21 48 58 & 0.1114 & 9 & 33396(\phn\phn\phn)
& 33224(\phn31) & 468(0) & 0.624(\phn87) & $<1.93$ & 11.14--11.49 \nl
19040+3356 & 19 04 04.1 & +33 56 12 & 0.1812 & 5 & 54322(300)
& 54151(302) & 787(5) & 0.715(\phn50) & 1.11(12) & 11.86 \nl
19084+3719 & 19 08 29.7 & +37 19 11 & 0.1091 & 2 & 32697(179)
& 32525(181) & 457(3) & 0.742(\phn67) & 2.81(31) & 11.59 \nl
19348+3400 & 19 34 50.4 & +34 00 04 & 0.1030 & 9 & 30889(\phn\phn\phn)
& 30684(\phn28) & 430(0) & 0.638(\phn38) & $<11.26$ & 11.08--11.98 \nl
19458+0944 & 19 45 52.0 & +09 44 31 & 0.1000 & 4 & 29964(\phn\phn9)
& 29729(\phn30) & 416(0) & 3.947(395) & 7.11(64) & 12.07 \nl
19559+1618 & 19 55 54.2 & +16 18 06 & 0.1396 & 9 & 41853(\phn\phn\phn)
& 41607(\phn27) & 593(0) & 0.880(\phn70) & $<2.26$ & 11.50--11.80 \nl
20318+2343 & 20 31 52.5 & +23 43 21 & 0.1011 & 9 & 30302(\phn\phn\phn)
& 30025(\phn23) & 420(0) & 0.944(104) & $<2.96$ & 11.23--11.58 \nl
20322+1849 & 20 32 14.1 & +18 49 45 & 0.1069 & 2 & 32038(111)
& 31756(113) & 446(2) & 0.749(\phn67) & 1.40(31) & 11.42 \nl
20344+0619 & 20 34 26.2 & +06 19 45 & 0.1645 & 2 & 49306(116)
& 49017(118) & 707(2) & 0.768(115) & 1.59(24) & 11.85 \nl
20361+1216 & 20 36 09.7 & +12 16 51 & 0.1320 & 2 & 39575(118)
& 39285(120) & 558(2) & 0.725(\phn65) & 0.94(\phn9) & 11.54 \nl
20394+2302 & 20 39 26.4 & +23 02 12 & 0.1053 & 2 & 31568(300)
& 31283(301) & 439(4) & 0.925(\phn65) & $<1.92$ & 11.26--11.52 \nl
20398+2745 & 20 39 53.4 & +27 45 28 & 0.1025 & 9 & 30715(\phn\phn\phn)
& 30437(\phn22) & 426(0) & 1.434(115) & $<8.28$ & 11.43--11.94 \nl
20402+1642 & 20 40 12.5 & +16 42 27 & 0.1378 & 2 & 41313(106)
& 41021(108) & 584(2) & 0.952(\phn86) & 1.16(\phn9) & 11.69 \nl
20450+0913 & 20 45 05.3 & +09 13 19 & 0.1218 & 2 & 36520(123)
& 36221(125) & 512(2) & 0.678(\phn54) & 1.43(17) & 11.52 \nl
20460+1925 & 20 46 01.8 & +19 25 49 & 0.1810 & 14 & 54262(300)
& 53967(301) & 784(5) & 0.883(\phn62) & $<1.45$ & 11.74--11.96 \nl
21026+1042 & 21 02 39.9 & +10 42 04 & 0.1078 & 2 & 32316(115)
& 32002(117) & 449(2) & 0.578(\phn52) & $<1.48$ & 11.08--11.38 \nl
21064+2155 & 21 06 29.4 & +21 55 35 & 0.1076 & 2 & 32257(300)
& 31949(301) & 449(4) & 0.598(\phn48) & 1.40(13) & 11.37 \nl
21135+0553 & 21 13 30.0 & +05 52 55 & 0.1058 & 2 & 31732(120)
& 31409(121) & 441(2) & 0.744(\phn67) & $<2.04$ & 11.17--11.48 \nl
21167+0819 & 21 16 42.7 & +08 18 57 & 0.1015 & 2 & 30443(138)
& 30117(139) & 422(2) & 0.568(\phn51) & 1.31(30) & 11.29 \nl
21251+1114 & 21 25 11.7 & +11 14 46 & 0.1140 & 2 & 34190(317)
& 33859(317) & 477(5) & 0.598(\phn54) & 0.79(\phn9) & 11.32 \nl
21256+0219 & 21 25 40.3 & +02 19 25 & 0.2570 & 16 & 77047(300)
& 76716(300) & 1154(5) & 0.285(\phn43) & $<0.65$ & 11.59--11.86 \nl
21329+0705 & 21 32 56.0 & +07 05 52 & 0.1165 & 5 & 34926(300)
& 34589(300) & 488(4) & 0.696(\phn56) & 0.85(16) & 11.40 \nl
21444+3534 & 21 44 25.3 & +35 34 34 & 0.1518 & 9 & 45523(\phn\phn\phn)
& 45220(\phn19) & 648(0) & 0.913(100) & $<1.93$ & 11.59--11.85 \nl
22057+0739 & 22 05 46.7 & +07 39 25 & 0.1178 & 2 & 35315(300)
& 34960(300) & 493(4) & 0.630(151) & 1.27(23) & 11.45 \nl
22068+2703 & 22 06 49.6 & +27 02 59 & 0.1550 & 5 & 46467(\phn\phn\phn)
& 46133(\phn14) & 662(0) & 0.640(\phn83) & 0.82(22) & 11.63 \nl
22416+1621 & 22 41 38.8 & +16 21 11 & 0.1945 & 2 & 58321(104)
& 57960(104) & 848(2) & 0.587(\phn59) & $<1.66$ & 11.63--11.95 \nl
22428+3215 & 22 42 50.7 & +32 15 44 & 0.1572 & 2 & 47130(114)
& 46799(115) & 673(2) & 0.652(\phn65) & $<3.23$ & 11.48--11.94 \nl
22541+0833 & 22 54 12.8 & +08 33 27 & 0.1661 & 13 & 49799(\phn42)
& 49431(\phn42) & 713(1) & 1.200(192) & 1.48(27) & 11.96 \nl
\enddata
\tablerefs{Redshifts were obtained from:  (1) \cite{kim98}; (2) W. Saunders
1999, personal communication;
(3) \cite{bee95}; (4) \cite{dow93}; (5) \cite{lee94}; (6) \cite{hil88}; 
(7) \cite{van95}; (8) \cite{str92}; 9) \cite{nak97}; (10) \cite{str88}; 
(11) \cite{kim95}; (12) \cite{sch83}; (13) \cite{fis95}; (14) \cite{vad93}; 
(15) \cite{deg92}; (16) \cite{ssg94}; (17) \cite{all85}.}
\end{deluxetable}

\begin{deluxetable}{cccccrrrrc} 
\scriptsize
\tablecaption{OH Non-Detections:  OH Limits and 1.4 GHz Properties\label{nondetectOH}}
\tablewidth{0pt}
\tablehead{
\colhead{{\it IRAS} Name} &  \colhead{$z_\odot$}  
& \colhead{$\log L_{FIR}$} & \colhead{$\log L^{pred}_{OH}$} 
& \colhead{$\log L^{max}_{OH}$} & \colhead{$t_{on}$} & \colhead{RMS} 
& \colhead{$f_{1.4GHz}$\tablenotemark{a}} & \colhead{Class} & \colhead{Note} \nl
\colhead{FSC} &  & \colhead{$h^{-2}_{75} L_\odot$} &
\colhead{$h^{-2}_{75} L_\odot$} & \colhead{$h^{-2}_{75} L_\odot$} & \colhead{min} 
& \colhead{mJy} & \colhead{mJy} & \nl
\colhead{(1)}& \colhead{(2)}& \colhead{(3)}& \colhead{(4)}& \colhead{(5)}& 
\colhead{(6)}& \colhead{(7)}& \colhead{(8)}& \colhead{(9)}& \colhead{(10)}
 }
\startdata
 02290+3139 & 0.2115 & 11.99 & 2.53 & 2.11 & 20 & 0.46 & 3.6(0.5) & & 4 \nl
 03477+2611 & 0.1494 & 11.71 & 2.15 & 2.03 & 8  & 0.76 & 4.1(0.5) & & 2 \nl
 03533+2606 & 0.1883 & 11.46--12.05 & 1.79--2.61 & 1.85 & 40 & 0.32 & 6.7(0.5) & & 1,4 \nl
 04046+1011 & 0.1845 & 11.50--12.16 & 1.85--2.76 & 2.05 & 16 & 0.52 & $<5.0$\phn & & 1\nl
 08559+1053 & 0.1480 & 11.89 & 2.39 & 1.69 & 20 & 0.35 & $<5.0$\phn & S2(1) &  \nl
 13542+1040 & 0.1234 & 11.47 & 1.81 & 1.67 & 12 & 0.48 & 65.1(2.4) & \nl
 14202+2615 & 0.1590 & 12.04 & 2.60 & 1.96 & 8 & 0.57 & 10.2(0.6) & H(2) & \nl
 14203+3005 & 0.1141 & 11.56 & 1.93 & 1.68 & 8 & 0.58 & 18.9(1.0) & S1.9(6)  \nl
 15445+3312 & 0.1558 & 11.52 & 1.88 & 2.12 & 4 & 0.85 & $<5.0$\phn & L(7)& 1 \nl
 15543+3013 & 0.1213 & 11.25 & 1.50 & 1.85 & 4 & 0.77 & $<5.0$\phn & H(7) 
& 1,2  \nl
 15597+3133 & 0.1437 & 11.54 & 1.91 & 1.88 & 8 & 0.59 & $<5.0$\phn &  &  \nl
 16045+2733 & 0.1139 & 11.56 & 1.93 & 1.61 & 12 & 0.50 & 9.7(0.6) &  \nl
 16121+2611 & 0.1310 & 10.79--11.05 & 0.86--1.23 & 1.94 & 4 & 0.80 & 18.2(0.7) & S1.5(8) & 1,6 \nl
 16156+0146 & 0.1320 & 11.69 & 2.12 & 1.82 & 8 & 0.59 & 8.6(0.5) & S2(9) &  \nl
 16284+2817 & 0.0970 & 11.39 & 1.70 & 1.46 & 16 & 0.44 & 4.2(0.5) &  \nl
 16474+3430 & 0.1115 & 11.88 & 2.38 & 1.66 & 8 & 0.59 & 11.5(0.6) &H(2)&  \nl
 17030+0457 & 0.1190 & 11.19--11.53 & 1.43--1.90 & 1.63 & 12 & 0.48 & $<5.0$\phn & S2(1)& 1 \nl
 17156+1238 & 0.1134 & 11.26--11.47 & 1.51--1.81 & 1.81 & 4 & 0.80 & 4.8(0.5) & H(3)&1,3  \nl
 17490+2659 & 0.1453 & 11.27--11.60 & 1.53--1.98 & 1.83 & 24 & 0.51 & 141.8(4.3) & S1(4) & 1,4  \nl
 17574+0629 & 0.1096 & 11.65--11.96 & 2.06--2.48 & 1.48 & 16 & 0.40 & 14.6(0.7) & H(2) &  \nl
 18030+0705 & 0.1458 & 12.00 & 2.55 & 1.85 & 12 & 0.53 & $<5.0$\phn &  \nl
 18040+2141 & 0.1016 & 11.64 & 2.05 & 1.42 & 24 & 0.41 & 5.7(0.5) &   \nl
 18147+1553 & 0.1024 & 11.59 & 1.97 & 1.45 & 16 & 0.43 & 53.4(1.7) &  \nl
 18222+1440 & 0.1262 & 11.47--11.71 & 1.81--2.14 & 1.75 & 8 & 0.57 & $<5.0$\phn &  \nl
 18315+2249 & 0.1310 & 11.44--11.72 & 1.76--2.16 & 1.68 & 12 & 0.45 & 16.0(0.7) & & \nl
18585+2148 & 0.1114 & 11.14--11.49 & 1.36--1.83 & 1.70 & 8 & 0.64 & 4.5(0.7) & & 1 \nl
19040+3356 & 0.1812 & 11.86 & 2.35 & 2.00 & 20 & 0.48 & 3.1(0.5) &  \nl
19084+3719 & 0.1091 & 11.59 & 1.98 & 1.57 & 16 & 0.50 & $<5.0$\phn & &  \nl
19348+3400 & 0.1030 & 11.08--11.98 & 1.27--2.51 & 1.49 & 12 & 0.47 & 3.3(0.6) 
& & 1 \nl
19458+0944 & 0.1000 & 12.07 & 2.64 & 1.53 & 8 & 0.55 & 15.3(1.0) & & 2   \nl
19559+1618 & 0.1396 & 11.50--11.80 & 1.85--2.26 & 1.78 & 16 & 0.50 & 50.8(2.2) & &  \nl
20318+2343 & 0.1011 & 11.23--11.58 & 1.48--1.96 & 1.49 & 16 & 0.49 & 3.3(0.5) & & 1 \nl
20322+1849 & 0.1069 & 11.42 & 1.74 & 1.76 & 12 & 0.81 & 3.3(0.6) & & 1 \nl
20344+0619 & 0.1645 & 11.85 & 2.33 & 1.99 & 12 & 0.58 & 4.5(0.5) &  \nl
20361+1216 & 0.1320 & 11.54 & 1.90 & 1.77 & 12 & 0.54 & 3.0(0.6) &  \nl
 20394+2302 & 0.1053 & 11.26--11.52 & 1.52--1.87 & 1.34 & 32 & 0.32 & 6.6(0.5) &  \nl
20398+2745 & 0.1025 & 11.43--11.94 & 1.75--2.45 & 1.71 & 4 & 0.80 & 14.8(0.7) & &  \nl
20402+1642 & 0.1378 & 11.69 & 2.11 & 1.84 & 12 & 0.59 & 6.4(1.5) & &2 \nl
20450+0913 & 0.1218 & 11.52 & 1.88 & 1.84 & 12 & 0.76 & 9.8(1.0) &  \nl
 20460+1925 & 0.1810 & 11.74--11.96 & 2.19--2.48 & 2.12 & 12 & 0.64 & 18.9(0.7) & S2(5)& 5  \nl
21026+1042 & 0.1078 & 11.08--11.38 & 1.27--1.68 & 1.40 & 32 & 0.35 & 10.4(0.6) & & 1  \nl
21064+2155 & 0.1076 & 11.37 & 1.67 & 1.65 & 12 & 0.63 & 5.4(0.5) &  \nl
21135+0553 & 0.1058 & 11.17--11.48 & 1.39--1.83 & 1.60 & 12 & 0.58 & 10.7(1.1) & & 1 \nl
21167+0819 & 0.1015 & 11.29 & 1.56 & 1.41 & 24 & 0.40 & 7.8(0.6) & &  \nl
21251+1114 & 0.1140 & 11.32 & 1.60 & 1.50 & 32 & 0.39 & $<5.0$\phn &  \nl
21256+0219 & 0.2570 & 11.59--11.86 & 1.97--2.35 & 2.37 & 12 & 0.56 & 3.6(0.6) & & 1 \nl
21329+0705 & 0.1165 & 11.40 & 1.71 & 1.59 & 20 & 0.46 & 2.4(0.5) & &  \nl
21444+3534 & 0.1518 & 11.59--11.85 & 1.98--2.34 & 1.89 & 20 & 0.54 & 14.8(0.6) &  \nl
22057+0739 & 0.1178 & 11.45 & 1.77 & 1.66 & 12 & 0.54 & 3.1(0.6) &  \nl
22068+2703 & 0.1550 & 11.63 & 2.03 & 1.97 & 12 & 0.63 & 4.3(0.5) &  & 2 \nl
22416+1621 & 0.1945 & 11.63--11.95 & 2.04--2.48 & 2.04 & 20 & 0.46 & $<5.0$\phn &  \nl
22428+3215 & 0.1572 & 11.48--11.94 & 1.82--2.46 & 1.99 & 12 & 0.63 & 8.8(0.5) & & 1  \nl
22541+0833 & 0.1661 & 11.96 & 2.49 & 1.86 & 24 & 0.42 & 5.7(0.6) & S2(9) & \nl
\enddata
\tablenotetext{a}{1.4 GHz continuum fluxes are courtesy of the NRAO VLA Sky Survey 
(\cite{con98}).}
\tablerefs{Spectral classifications were obtained from:  (1) \cite{hil88}; 
(2) \cite{kim98b}; (3) \cite{els85}; (4) \cite{deg92}; (5) \cite{fro89};
(6) \cite{mor96};  (7) \cite{vei95}; (8) \cite{dah88}; (9) \cite{vei99}.}
\tablecomments{(1) Source needs more integration time, based on
$L^{pred}_{OH} < L^{max}_{OH}$; (2) Source needs more integration time, due
to a suggestive feature in the bandpass; (3) Galaxy pair --- both nuclei 
have spectral type \ion{H}{2} (Elston {\it et al.} 1985); (4) Standing waves in 
the bandpass; (5) Sey 2/obscured Sey 1 (\cite{fro89}), BLR (\cite{vei97}); 
(6) Optically variable QSO (\cite{mac87}).}
\end{deluxetable}

\begin{deluxetable}{rccccrrrrcc}
\scriptsize
\tablecaption{OH Detections:  Optical Redshifts and FIR Properties 
		\label{detectFIR}}
\tablewidth{0pt}
\tablehead{
\colhead{{\it IRAS} Name} &  \colhead{$\alpha$} & \colhead{$\delta$} & 
\colhead{$z_\odot$} & 
\colhead{Ref} &
\colhead{$v_\odot$} & 
\colhead{$v_{CMB}$} & 
\colhead{$D_L$} &
\colhead{$f_{60\mu m}$} &
\colhead{$f_{100\mu m}$} &
\colhead{$\log L_{FIR}$} 
 \nl
\colhead{FSC} & \colhead{B1950} & \colhead{B1950} &\colhead{} &
\colhead{z} & \colhead{km s\minusone} & \colhead{km s\minusone} & \colhead{$h^{-1}_{75}$Mpc} & 
\colhead{Jy} & \colhead{Jy} & \colhead{$h^{-2}_{75} L_\odot$} \nl
\colhead{(1)}&\colhead{(2)}&\colhead{(3)}&\colhead{(4)}&\colhead{(5)}&
\colhead{(6)}&\colhead{(7)}&\colhead{(8)}&\colhead{(9)}&\colhead{(10)}&
\colhead{(11)}
}
\startdata
06487+2208 & 06 48 45.1 & +22 08 06 & 0.1437 & 2 & 43080(300)
& 43206(302) & 618(5) & 2.070(166) & 2.36(26) & 12.07 \nl
16300+1558 & 16 30 05.6 & +15 58 02 & 0.2417 & 1 & 72467(\phn64)
& 72515(\phn73) & 1084(1) & 1.483(134) & 1.99(32) & 12.43 \nl
17539+2935 & 17 54 00.1 & +29 35 50 & 0.1085 & 1 & 32525(\phn58)
& 32441(\phn67) & 456(1) & 1.162(\phn58) & 1.36(19) & 11.56 \nl
18368+3549 & 18 36 49.5 & +35 49 36 & 0.1162 & 3 & 34825(\phn40)
& 34688(\phn51) & 489(1) & 2.233(134) & 3.83(27) & 11.96 \nl
18588+3517 & 18 58 52.4 & +35 17 04 & 0.1067 & 1 & 31973(\phn35)
& 31810(\phn46) & 447(1) & 1.474(103) & 1.75(33) & 11.64 \nl
20248+1734 & 20 24 52.3 & +17 34 24 & 0.1208 & 1 & 36219(\phn87)
& 35943(\phn90) & 508(1) & 0.743(\phn82) & 2.53(38) & 11.66 \nl
20286+1846 & 20 28 39.9 & +18 46 37 & 0.1347 & 5 & 40396(127)
& 40117(129) & 571(2) & 0.925(\phn74) & 2.25(16) & 11.78 \nl
20450+2140 & 20 45 00.1 & +21 40 03 & 0.1284 & 5 & 38480(111)
& 38189(113) & 542(2) & 0.725(\phn51) & 1.90(15) & 11.64 \nl
21077+3358 & 21 07 45.9 & +33 58 05 & 0.1764 & 5 & 52874(117)
& 52587(119) & 763(2) & 0.885(\phn88) & $<1.55$ & 11.72--11.95 \nl
21272+2514 & 21 27 15.1 & +25 14 39 & 0.1508 & 5 & 45208(120)
& 44890(121) & 643(2) & 1.075(118) & $<1.63$ & 11.66--11.86 \nl
22116+0437 & 22 11 38.6 & +04 37 29 & 0.1939 & 5 & 58144(118)
& 57787(118) & 845(2) & 0.916(\phn73) & $<1.03$ & 11.82--11.98 \nl
& & & & & & & & & & \nl
19154+2704\tablenotemark{a} & 19 15 29.7 & +27 04 32 & 0.0994 & 4 & 29792(\phn\phn\phn) & 29601(\phn30)
& 414(0)& 1.502(120) & 2.85(23) & 11.66  \nl
\enddata
\tablenotetext{a}{{\it IRAS} 19154+2704 is an OH absorber.}
\tablerefs{Redshifts were obtained from:  (1) \cite{fis95}; (2) \cite{lu95}; (3) \cite{str92}; 
(4) \cite{nak97}; (5) W. Saunders 1999, personal 
communication.}
\end{deluxetable}

\begin{deluxetable}{cccccccrcccr}
\scriptsize
\tablecaption{OH Detections:  OH Line and 1.4 GHz Continuum Properties \label{detectOH}}
\tablewidth{0pt}
\tablehead{
\colhead{{\it IRAS} Name} &  
\colhead{$v_{1667,\odot}$} & 
\colhead{$t_{on}$} & 
\colhead{$f_{1667}$} & 
\colhead{$W_{1667}$} &
\colhead{$\Delta \nu_{1667}$\tablenotemark{a}} & 
\colhead{$\Delta v_{1667}$\tablenotemark{b}} & 
\colhead{$R_H$} & 
\colhead{$\log L_{FIR}$} & 
\colhead{$\log L^{pred}_{OH}$} & 
\colhead{$\log L_{OH}$} &
\colhead{$f_{1.4GHz}$\tablenotemark{c}}  \nl
\colhead{FSC} &\colhead{km s\minusone} & \colhead{min} 
& \colhead{mJy} & \colhead{MHz}
& \colhead{MHz} & \colhead{km s\minusone} &\colhead{} & 
\colhead{$h^{-2}_{75} L_\odot$} 
& \colhead{$h^{-2}_{75} L_\odot$} & \colhead{$h^{-2}_{75} L_\odot$} & 
\colhead{mJy} \nl
\colhead{(1)}&\colhead{(2)}&\colhead{(3)}&\colhead{(4)}&\colhead{(5)}&
\colhead{(6)}&\colhead{(7)}&\colhead{(8)}&\colhead{(9)}&\colhead{(10)}&
\colhead{(11)}&\colhead{(12)}
}
\startdata
06487+2208 & 43017(12) & 28 & \phn7.60 & 0.85 & 1.03 &    211 & 6.1 & 12.07 & 2.63 & 2.86 & 10.8(0.6) \nl
16300+1558 & 72528(12) & 16 & \phn3.12 & 0.56 & 0.59 &    131 & \nodata & 12.43 & 3.14 & 2.81 & 7.9(0.5) \nl
17539+2935 & 32522(12) & 80 & \phn0.76 & 0.72 & 0.81 &    161 & $\geq 2.9$ & 11.56 & 1.93 & 1.74 & 4.0(0.6) \nl
18368+3549 & 34832(12) & 32 & \phn4.58 & 1.79 & 2.10 &    421 & $\sim$ 9.5 & 11.96 & 2.48 & 2.83 & 21.0(0.8)\nl
18588+3517 & 31686(12) & 32 & \phn7.37 & 0.56 & 0.32 & \phn64 & 5.1 & 11.64 & 2.05 &  2.50 & 5.9(0.5) \nl
20248+1734 & 36538(12) & 48 & \phn2.61 & 1.36 & 0.88 &    177 & $\sim$ 6.8 & 11.66 & 2.07 & 2.51 & $<5.0$\phn \nl
20286+1846 & 40471(12) & 24 &    15.58 & 1.51 & 1.10 &    224 & $\geq 4.4$ & 11.78 & 2.23 & 3.38 & $<5.0$\phn\nl
20450+2140 & 38398(12) & 44 & \phn2.27 & 0.67 & 0.71 &    144 & $\geq 6.2$ & 11.64 & 2.05 & 2.21 & 5.0(0.5) \nl
21077+3358 & 52987(12) & 28 & \phn5.04 & 1.86 & 1.15 &    243 & $\geq 7.4$ & 11.72--11.95 & 2.16--2.47 & 3.23 
& 9.4(1.0) \nl
21272+2514 & 45032(12) & 32 &    16.33 & 1.87 & 1.27 &    263 & 13.7 & 11.66--11.86 & 2.07--2.34 & 3.63 
& 4.4(0.5) \nl
22116+0437 & 58180(12) & 68 & \phn1.76 & 1.16 & 0.56 & 121 & $\sim$ 5.2 & 11.82--11.98 & 2.30--2.52 & 2.74 
& 8.4(0.6) \nl
& & & & & & & & &   \nl
19154+2704\tablenotemark{d} & 29894(12) & 32 & -2.62 & 0.85 & 0.93 & 184 & 1.81 & 11.66 & 2.07 
& \nodata & 63.6(2.0)\nl
\enddata
\tablenotetext{a}{$\Delta \nu_{1667}$ is the {\it observed} FWHM.}
\tablenotetext{b}{$\Delta v_{1667}$ is the {\it rest frame} FWHM.  The rest
frame and observed widths are related by 
$\Delta v_{rest} = c(1+z)(\Delta\nu_{obs}/\nu_\circ)$.}
\tablenotetext{c}{1.4 GHz continuum fluxes are courtesy of the NRAO VLA Sky 
Survey (\cite{con98}).}
\tablenotetext{d}{{\it IRAS} 19154+2704 is an OH absorber.}
\end{deluxetable}

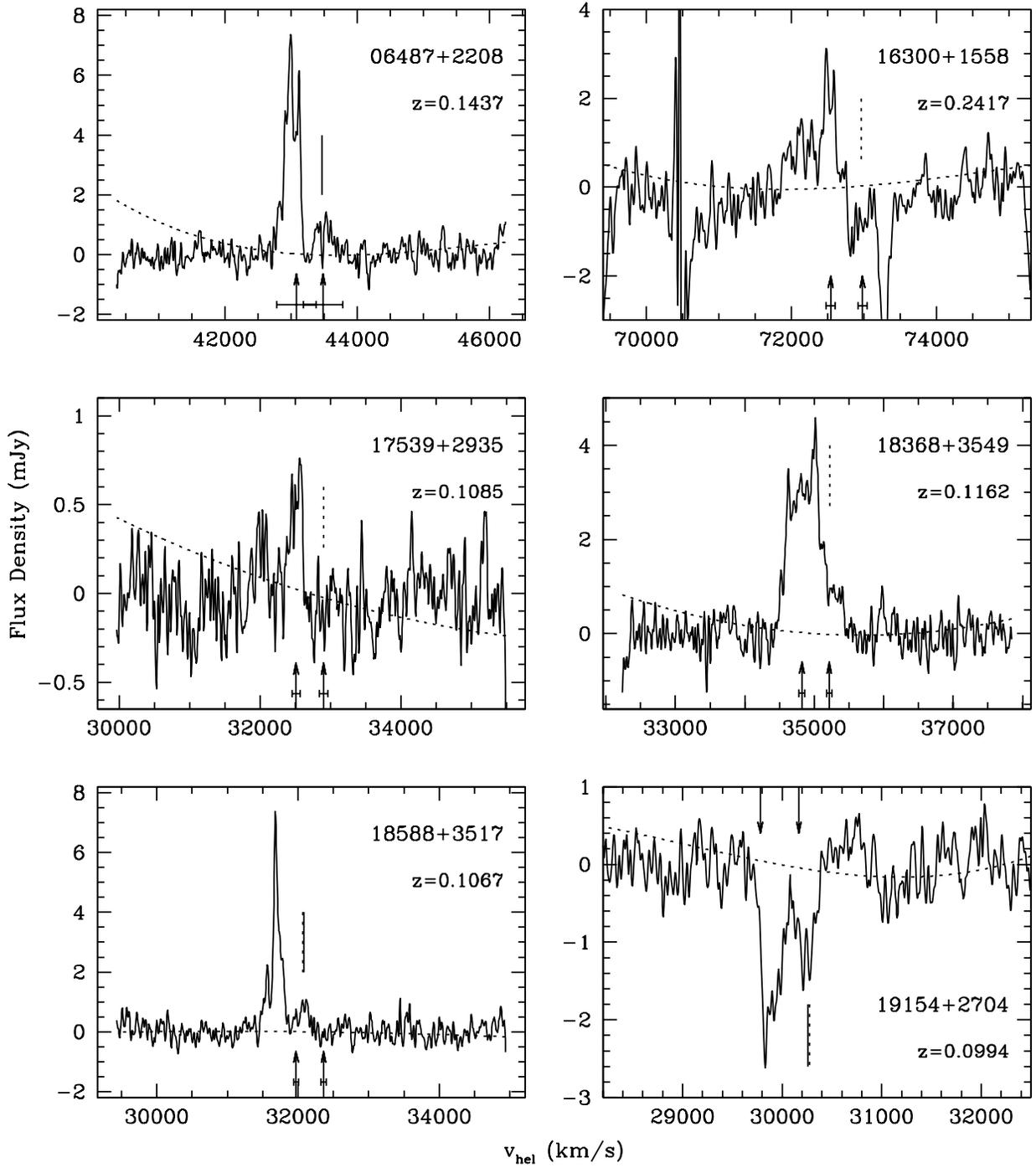
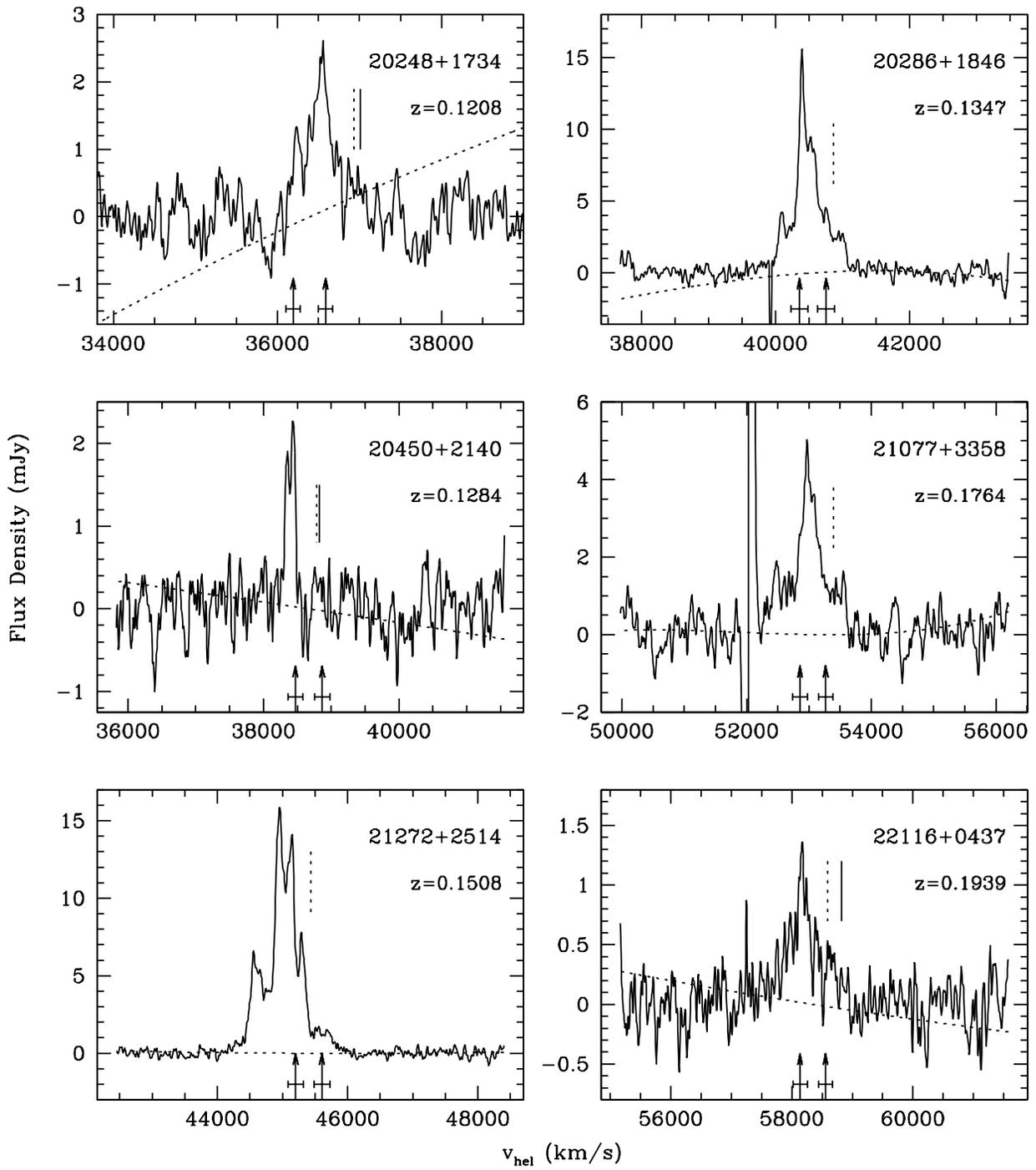
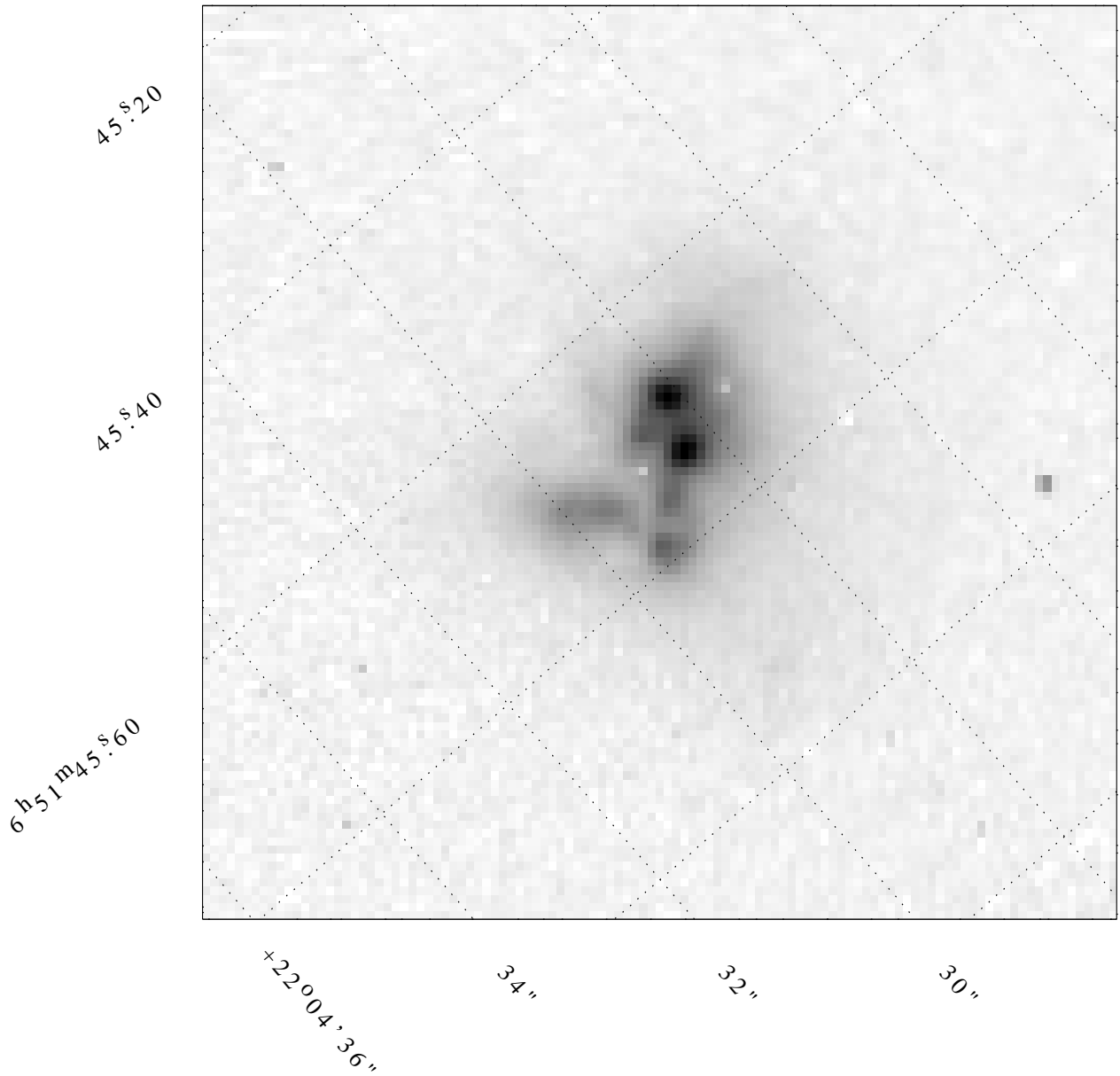
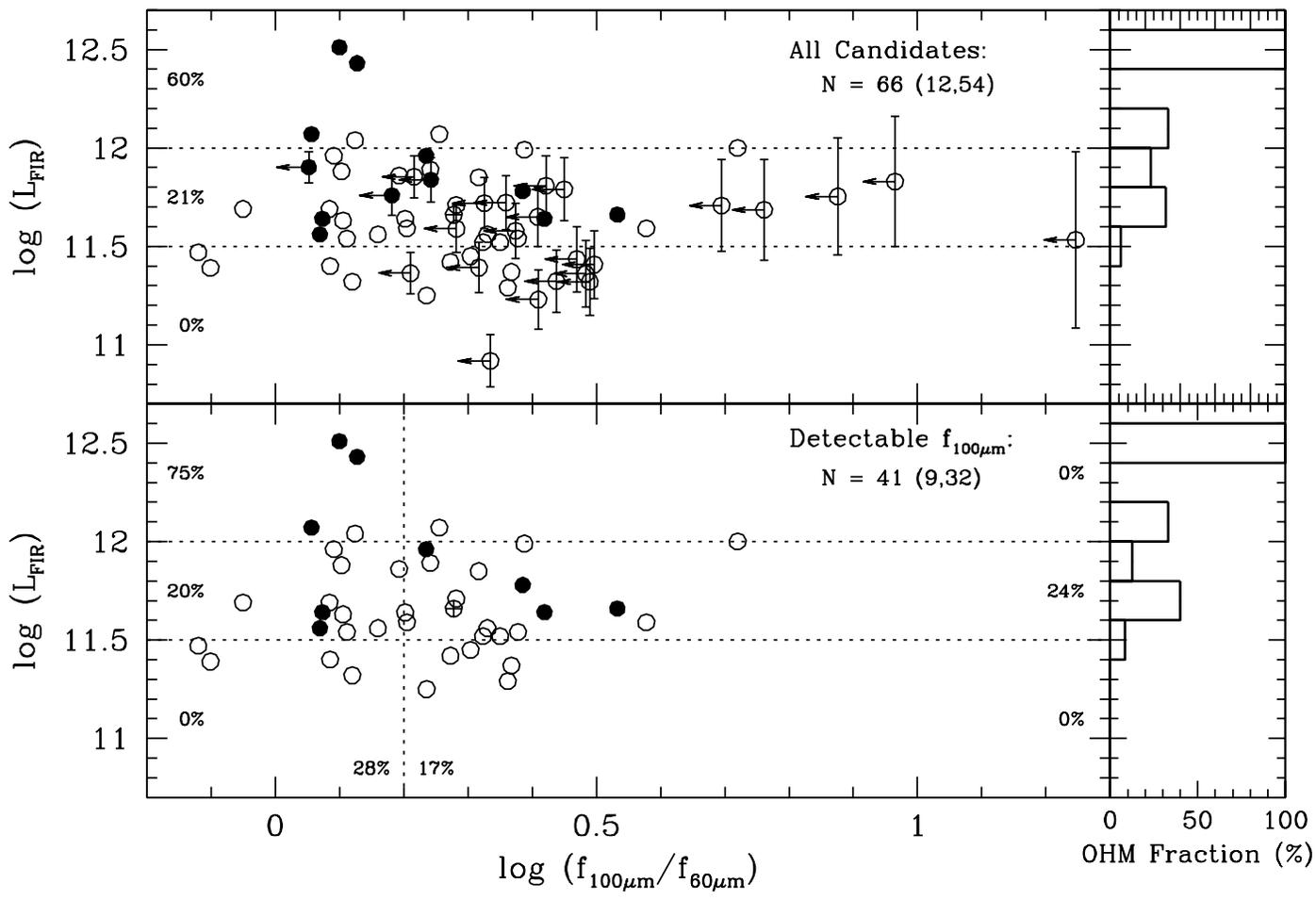
\end{document}